\documentclass[a4paper]{article}
\usepackage{graphicx}
\newcommand{\be}{\begin{equation}}
\newcommand{\ee}{\end{equation}}
\newcommand{\bd}{\begin{displaymath}}
\newcommand{\ed}{\end{displaymath}}
\newcommand{\bea}{\begin{eqnarray}}
\newcommand{\eea}{\end{eqnarray}}
\newcommand{\bi}{\begin{description}}
\newcommand{\ei}{\end{description}}
\newcommand{\bq}{\begin{quote}}
\newcommand{\eq}{\end{quote}}

\def\fo{\footnote}

\def\a{\alpha}
\def\b{\beta}

\def\D{\Delta}
\def\e{\epsilon}

\def\om{\omega}
\def\Om{\Omega}
\def\l{\lambda}

\def\s{\sigma}

\def\apj{Astrophysical Journal}

\begin{document}
\bibliographystyle{unsrt}

\author{Alexander~Unzicker\\
        Pestalozzi-Gymnasium  M\"unchen\\[0.6ex]
{\small{\bf e-mail:}  alexander.unzicker et lrz.uni-muenchen.de}}

\title{Why do we Still Believe in Newton's Law~?\\
Facts, Myths and Methods in Gravitational Physics}
\maketitle

\begin{abstract}
An overview of the experimental and observational    
status in gravitational physics is given, both for the known tests of general relativity
and Newtonian gravity, but also for the increasing number of results where
these theories run into problems, such as for dark matter, dark energy,
and the Pioneer and flyby anomalies. It is argued that (1) scientific
theories should be tested (2) current theories of gravity
are poorly tested in the weak-acceleration regime (3) the measurements suggest that the
anomalous phenomena have a common origin (4) it is useful to consider
the present situation under a historical
perspective and (5) it could well be that we still do not understand gravity.
Proposals for improving the current use of scientific methods are given.


\bq
`We do not know anything - this is the first. Therefore, we should be very modest - this is the second.
Not to claim  that we do know when we do not- this is the third. That's the kind of attitude
I'd like to popularize. There is little hope for success.' (Karl Popper)
\eq

\end{abstract}

\section{Introduction}
For by far the longest part of history
gravity was just the qualitative observation that earth attracts objects, and for about 1500 years,
this was a strong argument backing
the geocentric model of Ptolemy. 
Its accurate, but complicated epicycles with excentrics, equants
and deferents had hidden the better and simpler ideas thought
already by the Greek astronomer Aristarchus. King Alfonso X of
Spain, who learned the Ptolemaic system from Arab libraries in his
country \cite{Sin},
 commented the system of epicycles:
`If the Lord Almighty had consulted me before embarking upon
Creation, I should have recommended something simpler.' Though
this seems common sense, the Copernican 
revolution, backed by the
excellent observations of Tycho Brahe, by Kepler's ingenious
descriptive laws and by Galilei's famous first use of the
telescope, was a difficult process until it terminated with the
general acceptance of Newton's law of gravitation. What lasted
that long for the human perception was a blink of an eye with
respect to the age of the universe, of which we now can take the
Hubble time $H_0^{-1} \approx 14$ billion years as a measure.

Astronomy became modern Astrophysics and Cosmology, and for the last two decades mankind
 collected data of unique precision. Satellite-based telescopes covering all frequencies
and digital image
processing were not a minor revolution than the application of the telescope itself
in 1608. Theoretically, this enormous amount of surprising data is described by a
model of standard cosmology. This, in the light of the history of
gravitation, fast digestion of data is accompanied by an increasing number of
parameters $H_0, \Om, \Om_b, \Om_{DM}, q_0, \Lambda, \tau, A_s, n_s$\fo{Hubble constant, density parameter,
baryonic density, density of dark matter, deceleration parameter, cosmological constant, optical depth
scalar fluctuation amplitude, scalar spectral index. See also,
 e.g. \\ $http://en.wikipedia.org/wiki/Lambda-CDM\_model$.}.
While the numerical values become more and more precise,
little advance has been achieved in the question as to what the related quantities
`dark matter' and `dark energy' do really mean. A soulmate
of King Alfonso,
the astronomer Aguirre \cite{Agu:01a} commented: `these new
discoveries...  have been achieved at the expense of simplicity'. 
The question why nature comes up with a bunch of numbers like
$0.73$ must be allowed or at least not forgotten. Though standard cosmology has  various
free parameters now, it had started from a profound but simple physical concept, the equivalence
principle. Based on that, in 1915 Einstein revealed  a
deep relation of gravity to the geometry of spacetime, which was enthusiastically
accepted
after the verification of light deflection in 1919 and the explanation
of the anomalous advance of the perihelion of mercury known since 1859.
From the 1960s until today, general relativity (GR) has
undergone an impressive series of confirmations I will briefly review below. However,
the following comment in a textbook of galactic astronomy (\cite{BinGD}, p.~635),
is remarkable:
\bq
`It is worth remembering
that all of the discussion so far has been based on the premise that
Newtonian gravity and general relativity are correct on large scales.
In fact, there is little or no direct evidence that conventional
theories of gravity are correct on scales much larger than a parsec
or so. Newtonian gravity works extremely well on scales of
$\sim 10^{14} cm$ (the solar system). (...)  It is principally the
elegance of general relativity and its success in solar system
tests that lead us to the bold extrapolation that the gravitational
interaction has the form $GM/r^2$ on the scales $10^{21}-10^{26} cm$...'
\eq

While the tests of GR
mostly regard {\em strong\/} fields, the phenomena which actually are not explained yet like
galaxy rotation curves (dark matter), the Pioneer anomaly etc. seem to occur in {\em weak\/}
fields where GR does not distinguish from the Newtonian limit.
The  above cited  bold extrapolation  seems to have encountered
observational problems even in the solar system \cite{Lam:06} now.

The experimental situation motivates to try an overview on all possible and 
available tests of gravity.\fo{The growing interest is also reflected by the
announced review \cite{Wil:08}.}
Due to the amount of material, I  must refer to other
review articles where appropriate. The experimental and observational
evidence is presented at various distances, masses, and accelerations. This
last ordering is particulary adopted to put into evidence the big gaps where
our  knowledge is not really tested.

The major part of researchers will be happy to fine-tune
the standard model
with better data. This article instead is addressed to those who think
we have gone to the limits in introducing new parameters \cite{Sha:05}
in gravitational physics, and instead of the excentrics and deferents 
adjusting the epicycles, it may be time to think about alternatives.

While presenting alternatives is not the task of this article, we shall
touch some historically interesting attempts briefly. On the other hand,
speculations on extensions or modifications of gravity have to match a huge
amount of high-quality data. Since it is not easy to collect results from very different
research areas, the overview given below can be a first guide
to facts that must be taken into consideration. 

\section{Observational and experimental tests}
\bq
{\it `Theories crumble, but good observations never fade' - Howard Shapley}
\eq

\paragraph{Newton's law of gravitation}
\be F= G \frac{M m}{r^2}, \label{newton} \ee in which I shall call
$M$ field mass and $m$ test mass, must be tested
under several aspects. The dependence on $m$, in particular the
material {\em in\/}dependence of the acceleration $G
\frac{M}{r^2}$ is usually
called (weak) equivalence principle. 
Many tests refer to the so-called inverse-square law in distance, but we will also
discuss observations that test the dependence on $M$. Since mass estimates of the whole
universe depend on it, determining the absolute value of $G$ is very important, and
fortunately, a lot of research has been done in the past years here. There are however
independent tests of the spatial and temporal dependence of $G$, which can be
seen as tests of Newton's law as well.

\paragraph{Einstein's theory of general relativity,} at the time of its
development,  had only a few observational confirmations. This changed in the 1960s \cite{Wil}, 
and a lot of spectacular measurements were in perfect agreement with GR. Since
the recent focus of interest are tests in the low-acceleration-scale where
GR is Newtonian, I shall briefly mention GR tests and refer the reader to specific
reviews \cite{Ber:06, Wil:05, Wil:06, Tur:08}. 

\subsection{Tests ordered by distance}
The tests described in the following involve measurements of $G$, its possible
variation, tests of the $1/r^2$-dependence and of the equivalence principle.
The absolute $G$ measurements are unique in the sense that the mass determination
of the whole universe rely on them; all other measurements yield mass ratios.

\paragraph{Atomic scale.}
Though it is common folklore to calculate the ratio of electric
and gravitational forces ($m_p, m_e$ proton and electron mass) as
\be \frac{F_e}{F_g} = \frac{e^2}{4 \pi \e_0 G m_e m_p} = 2.27 \
\times \  10^{39}, \ee this number has never been measured, since
elementary particles are too light for being used as field masses.
With the earth as field mass instead, $G$ measurements with atom interferometry
have reached a remarkable accuracy \cite{Fix:07, Lam:08}.

\paragraph{Sub-millimeter-scale.}

A violation of the inverse-square-law was suggested by fifth force
and string theories. The E\"ot-Wash group \cite{Ade:03, Hoy:04, Kap:07},
using a sophisticated version of the torsion balance, obtained 
good evidence for the validity of the inverse square law for
masses separated on the above scale and obtained tight constraints
on the violation parameters $\a$ and $\l$. These are difficult
experiments, since electrostatic effects and the Casimir effect
\cite{Lam:97,Bre:02} have to be eliminated carefully. The claim of
a sub-mm-test is however a little misleading, since the
barycentric 
distance of the masses was much greater.

\subsubsection{Laboratory scale.}
\paragraph{Torsion balance.}
The most famous, important, and precise measurement
of the gravitational force uses the 200 year old Cavendish torsion balance,
to which only minor modifications were applied until recently. The distance is usually
$10-20 \ cm$.
Long \cite{Lon:76} claimed to have observed deviations of the inverse-square law
at laboratory distances. It seems however that these experiments have not
received independent confirmation.
About 20 years
ago, the determination of $G =6.6726 \times 10^{-11} m^3 s^{-2} kg^{-1}$
 by \cite{LuT}
was considered to have an uncertainty of 0.013 \%. In 1995, the
PTB \cite{Mic:95} published a much higher, probably wrong value
for $G$, but this paper encouraged many groups to perform new
measurements of $G$. In the following, astonishing discrepancies
arose and led to an increase of the uncertainty in the CODATA
value of $G$ to 0.15 \%. The 1982 measurement  \cite{LuT} was
shown to bear systematic errors \cite{Kur:97}, which were
corrected much later \cite{Bag:97} to $G=6.6740 \pm
0.0007$.\fo{For simplicity, the unit  $10^{-11} m^3 s^{-2}
kg^{-1}$ will be dropped in the following} A different setup with
a rotating torsion balance was used by recent precision
measurements \cite{Gun:00} and \cite{Qui:01}. Though their values
$G =6.674215 \pm 0.000092... $ and $G =6.67559 \pm 0.00027$ are
still more discrepant than the respective error bars, the
controversy seems to be settled.

\paragraph{Gravitational redshift.} With the recoil-free
emission of X-ray
photons from crystals (M\"ossbauer effect), frequencies could be measured
with an accuracy unknown so far. Pound and Rebka \cite{Pou:60,Pou:65} used
this to demonstrate the gravitational redshift of a photon leaving the field
of the earth. This is the only test of GR at the laboratory scale.

\subsubsection{Movable field masses - intermediate scale ($1-100 \ m$)}
There are few other laboratory methods.
\cite{Bal:05} measured the effect of a moving mass of 280 $kg$ on a
superconducting gravimeter an obtained $G=6.675 \pm 0.007 $. 
A similar setup, though with a `free fall' method was used by \cite{Sch:99} with the
result $G =6.6873 \pm 0.0094$ and a field mass of $500 \ kg$.
\cite{Sch:02} used a beam balance with $13521 \ kg$ of mercury as field masses
Their measurement $G =6.67407(22)$ at a distance of $1 \ m$ agreed with
 other recent values (see the references in there).
The above superconducting gravimeter was  used by \cite{Bal:01} to determine $G$
from the water level variation of a little storage lake, with the result $G=6.688 \pm 0.011$.
\cite{Hub:95} obtained $G=6.678 \pm 0.007$ and $G=6.669 \pm 0.005$ with a lake experiment
using a precision balance instead of a gravimeter. The effective distances were $88 \ m$ and $112 \ m$,
respectively, and this much more accurate than the value $G=6.689 \pm 0.057$ obtained in
an early lake experiment with effective distance of $22 \ m$ \cite{Moo:88}.
A review on early intermediate scale measurements is \cite{Sta:87}, whereas a review
on the discrepant values in the 1990s is \cite{Gil:99}.

To summarize, the very discrepant measurements of $G$ in the 1990s seem to converge to
a commonly accepted value of $G=6.674$. This accuracy still cannot compete with
other constants of nature that reach a relative precision of $10^{-12}$.
New ideas for space-based experiments were reviewed in \cite{San:96}.
On a meta-level, a result
 of the past decade is however that
precision measurements of $G$ are of enormous difficulty and
therefore many groups tended to overestimate the accuracy of their
results. Going into the details, in  many papers one can find
considerable variation in the single measurements which is
believed to be of statistical nature. This holds also for
\cite{Gun:00} and \cite{Sch:02}. On the other hand, no convincing
mechanism for a variation of $G$ could be backed by the
experiments.

\subsubsection{Geophysical scale}
A couple of interesting measurements with gravimeters have been
conducted by measuring the gradient of $g$, which can be
calculated if the density of the surrounding material is known.
While the measurements on  towers \cite{Tho:89, eck:88} were
disputed and not confirmed \cite{Rom:97},
 a discrepancy from Newton's law was found for a mine
hole \cite{Hol:86, Fis:91}. Interestingly, a discrepancy in the same
direction was found independently for a hole in the Greenland ice
cap \cite{And:89}, where good estimates for the ice density were
available. Despite the anomalous gravity gradient, they concluded,
however, `we cannot unambiguously attribute it to a breakdown of
Newtonian gravity because it might be due to unexpected geological
features in the rock below the ice.' This is a general problem of
all experiments with {\em moving\/} gravimeters - the uncertainty
of the density distribution of the earth crust usually limits
accuracy. In a similar submarine experiment, $G=6.677 \pm 0.013$
was obtained \cite{Zum:91}, in agreement with laboratory values.

\paragraph{Spatial variation of $G$.}
The GGP (global geodynamics project) network of superconducting gravimeters allows a
high-precision measurement ($10^{-12} \frac{m}{s^2}$) of local gravity $g$.
 Though no absolute values
can be obtained due to a drift (which is clearly instrumental, but of unknown origin yet),
variations due to tides, air pressure and a variety of geophysical effects can be monitored and modelled.
In principle, the slightly elliptic earth orbit allows also to test a possible
spatial dependence of $G$ \cite{Unz:06}.

\paragraph{E\"otv\"os experiments, equivalence principle.}

In 1907, the Hungarian baron showed the material-independence of the gravitational
acceleration $g$ with an extraordinary precision. Though unknown to Einstein, this
experiment confirmed the theoretical basis of GR, the equivalence principle (EP).
The accuracy was greatly improved by experiments in the 1970s \cite{Dic:64, Bra:72}.
An even greater accuracy could be achieved by satellite experiments in projection \cite{Com:06}.
See also \cite{Wil:05, Wil:06} for an overview.

\paragraph{The Hafele-Keating clock experiment.}
In 1972, two atomic clocks were transported in airplanes orbiting
the earth eastwards and westwards  \cite{HKe}.
 Besides the SR effect of moving clocks
that could be eliminated by the two flight directions, the results
showed confirmation of the first-order general
relativistic time delay, whose accuracy was improved
later \cite{Ves:80}.
\subsubsection{Satellite scale.}

\paragraph{Lunar Laser Ranging (LLR).}

There are no other experiments determining $G$ directly on larger
scales. The product $G M_{E}$ however can be measured by orbital
data of the moon and artificial satellites. LLR, which became
possible after the Apollo missions where reflectors were left, has
reached an extraordinary accuracy of some $cm$, and in the near
future it will go even below (see \cite{Wil:03}
for a review, and \cite{Bat:07, Mer:07}). It is therefore suited 
to put constraints on parameterized post-Newtonian (PPN)
parameters and $\dot G$. The problem however is that a change in
$G$ which results in a distance variation earth-moon, is masked by
a well-known distance increase due to the tidal friction. Earth
rotation is slowed down by energy dissipation, and the earth-moon
system has to conserve its angular momentum. Since the dissipation
process is very hard to quantify, an independent measurement of
the earths rotation slowdown is sought. 
Recent investigations
with ancient solar and lunar eclipses \cite{Ste:03} showed a
discrepancy of so far unknown origin. A similar claim is stated by
\cite{Dum:02}. A possible time dilation is claimed by \cite{Dei:07}.

\paragraph{LAGEOS.}
The LAGEOS satellites orbit earth at a height of 5900 km and, so
to speak, consist of mirrors only. This `cannonball' type of
satellite was designed with a minimum of disturbing components
thus allowing various precision measurements by analyzing its
orbital data. For instance, the general relativistic perigee shift
could be verified \cite{Ior:02}. Combining the earth gravity field
data of the CHAMP and GRACE satellites, an even better accuracy is
expected. Moreover, the launch of an improved satellite LARES is
planned \cite{Ciu:06}. Recently, LAGEOS data were used to test the
Lense-Thirring effect and geodetic precession \cite{Ciu:02}, which
should independently be tested in the near future by the gravity
probe B mission launched in 2004.

\paragraph{APSIS}
is an artificial planetary system in space designed to test gravity under
unpreceded conditions \cite{Sha:06}.

\paragraph{LATOR.}
The Laser Astrometric Test Of Relativity is a satellite-based
Michelson-Morley-type experiment that will use optical
interferometry by interplanetary laser ranging. The accuracy of
the GR parameters measured so far will be greatly improved and
further parameters will be determined which were never measured
before. The LATOR results will be able to distinguish various
extensions and modifications of GR \cite{Tur:05a}.

\paragraph{The flyby anomaly.}
The swing-by technique for satellites is used to change the
direction and heliocentric velocity of spacecraft \cite{And:06}.
In various occasions, after a swing-by process at the earth,
satellites showed an unexplained velocity increase $\D v$, to
which for years little attention was given. Until now, it has been
observed three times independently (Galileo, NEAR, Rosetta, see
\cite{And:06}), though under very different conditions and with a
great variation in the amount of $\D v$. Recently, a possible
dependence on the eccentricity and the perigee distance was
suspected \cite{Lam:06}. While the existence of the effect is quite
accepted, further data is needed for a systematic description of
this puzzling behavior. The hyperbolic trajectory of all
occurrences seems to be the main difference to many other
well-tested satellite orbits. With respect to the extensive data
on bound orbits, the observational basis of escape orbits is very poor
\cite{Lam:07}.

\subsubsection{Inner solar system scale.}

Planetary orbits allow to test the inverse-square-dependence or
the constancy of Kepler's constant $G M_{sun} =\frac{4 \pi
a^3}{T^2}$, where $a$ is the semimajor axis and $T$ the time of
revolution. The precision of planetary orbits is partly obtained
by astronomical observations, such as transits of mercury and
Venus. Distance measurements are obtained from the reflection of radar
signals at Venus, and in particular from the Viking lander
missions on Mars 1979-1982 \cite{Rae:79}. No deviations from
Kepler's law but constraints on $\dot G/G$ were found\fo{See
\cite{Uza} for a review on $\dot G/G$ constraints.}. The precise data allow 
even to constrain dark matter \cite{Ser:06b} and dark energy \cite{Ser:06a, jet:07}.

\paragraph{Classical tests of GR.}
Three of the four classical tests of GR take place at these
distances. The deflection of light passing near the sun, measured
for the first time in 1919, has now been measured to be within the
predicted value by $0.1 \%$ \cite{Wil}. The perihelion advance of
mercury, known since 1859, has shown less progress in precision
\cite{Cle:47, Sha:90}. The Shapiro time delay \cite{Sha:64} of radar
signals passing nearby the sun has been measured in the 1960s for
the first time. The best agreement with GR is currently obtained by
the Cassini spacecraft data \cite{Kop:06}.

\paragraph{Helioseismology,} the analysis of acoustic waves of the sun, has become
an interesting area of research. Relevant for gravitation are the constraints
on a possible variation of $G$ which reach $\dot G/G < 10^{-12} yr^{-1}$
\cite{Uza, Gue:98}.

\subsubsection{Outer solar system scale.}
Data from the outer Planets were collected by the numerous
satellite missions Pioneer 10 and 11, Galileo, Ulysses, Voyager
and Cassini. Among other important scientific results which are
not addressed here, radio tracking techniques improved the orbital
data accuracy \cite{Ior:06a}. While planetary data showed no hint
for a violation of Newton's law\footnote{with the recent exception 
\cite{Ior:08}}, a surprising anomalies were
observed regarding the motion of the spacecraft itself.

\paragraph{The Pioneer anomaly.}
Though Pioneer 10 and 11 were launched in 1972 and 1973 already,
the first detailed investigation of predicted and observed motion
was published in 1998 \cite{And:98}. This had mainly three
reasons: (1) nobody had expected a deviation, (2) the effect is
small and difficult to separate from other influences, (3) it
seemed to have occurred after the last maneuvers and planetary
flybys in 1974/79. The anomaly consists of an unmodelled 
acceleration $a_p= 8.74 \pm 1.33 $x$ 10^{-10} \frac{m}{s^2}$
directed towards the sun, or equivalently, an anomalous blue shift
drift $(2.92 \pm 0.44) $x$10^{-18} \frac{s}{s^2}$ of the radio
tracking signal \cite{And:01, Tur:05, Nie}. In the meantime, an
enormous effort has been conducted to model whatever physical
effect that could influence the spacecraft motion. Two  possible
explanations favored by group members were: (1) some effect
related to the heat produced by the spacecrafts energy source (2)
gas leaks that lead to an acceleration. Both hypotheses became
more and more unlikely, because (1) the decreasing heat production
should have translated into a decreased force that has not been
observed (2) gas leaks would require an astonishing constancy and
the unlikely coincidence to be aligned with motion for all
spacecrafts.

However, this remarkable observation would never have attracted so much
attention if it hadn't contained a challenge for theoreticians: the numerical 
coincidence of $a_p$ with $c H_0$, the product of speed of light and the Hubble constant. A review
of theoretical speculations explaining the Pioneer anomaly is given in \cite{And:01}.
Probably most of these proposals are going to be ruled out by  orbital data of the outer
planets, which have shown to be incompatible with an extra acceleration of the amount
$a_p$ \cite{Ior:06a, Ior:06c, Ior:06e}. Considerations on comets
and minor planets \cite{Pag:05, Wal:07} give hope that in the near future it can  be tested
if $a_p$ influences highly elliptic orbits.
While new missions are proposed to test the anomaly \cite{Nie:04, Nie:07a}, the analysis of
newly recovered data \cite{Tur:06} will be extremely interesting. Since the anomaly
of Pioneer 11 seems to have started with the last flyby (Saturn), the old data, containing
the Pioneer 10 Jupiter flyby, will reveal if there is a link to the flyby anomaly \cite{And:06}.
Besides the constant acceleration, there  are anomalous daily and annual signals, too.
Since there are more possibilities for systematic errors \cite{And:01},
less importance has been given to those.

\subsubsection{Extrasolar scale.}

Very little gravity tests are possible in the range just above the
solar system. This has changed a little with the discovery of
planets. Though the effective distance is of the same order as
the solar system (0.01-5 AU), Iorio \cite{Ior:06} deduced limits
on the spatial variation of $G$ using orbital data from {\em
www.exoplanet.eu\/}.

\subsubsection{Globular cluster scale.}
The first test of Newtonian gravity in the scale of $20-50 \
pc$\fo{$1 \ pc = 3.262$ light years.} relies on very recent
observations. Globular clusters, dominated by radial motion of
stars, cannot be accessed by observing rotation curves like
galaxies (see below). Following the suggestion of \cite{Bau:05},
 \cite{Sca:06, Sca:06a, Dru:07} investigated
velocity dispersion curves for $\om Cen$, $M \ 15$, $NGC \ 6171$, 
$NGC \ 6341$  and $NGC \ 7099$.  Instead of an expected Keplerian
falloff, the curves show a flattening of the velocity profile, 
at the same {\em acceleration $a_0$\/} that has been observed
at galactic rotation curves (see sec.~2.3.2).
Remarkably, this phenomenon has been confirmed at the low-concentration cluster
$NGC \ 288$ which has an entirely flat dispersion profile \cite{Sca:07}.

\subsubsection{Galactic scale.}

\paragraph{Galactic rotation curves} of about 1000 Galaxies
\cite{Per} have provided
the by far strongest evidence for the disagreement of 
`dynamical' and visible mass. Assuming that
all mass of a spiral galaxy is contained within its optical
radius, one expects due to \be v^2 = \frac{G M}{r} \ee a radial
dependency $v \sim r^{-\frac{1}{2}}$ in the velocity profile of
clouds that can be measured by Doppler shifts.
Interestingly, up to multiples of the optical radius
practically all galaxies show rather constant (`flat') velocities 
than the expected Keplerian behaviour.
Usually, an explanation with `dark matter' is given,
though this requires a particular distribution. While the deviation is 
already visible within the optical radius, in the outer regions ratios
of dark and luminous matter up to 1000 are required \cite{Sal:96}. 
The form of the galactic rotation curves seem to depend just on the size
of the galaxy \cite{Sal:96, Sal:07a}, a fact which is hard to explain
by the properties of any dark matter candidate. Many precision profiles
are conflict with the standard model \cite{Gen:04}, among these the most extended
velocity profile of NGC 3741 \cite{Sal:07}.
While many questions are still open \cite{Sal:02, Sal:03} 
the anomaly itself is beyond 
any experimental doubt (see overviews \cite{Agu:01a, Bos:98, Rub:01}).
There are clear hints that the morphology of galaxies is dominated by
systematics we do not understand yet \cite{Dis:08, vBe:08}.

\paragraph{Low surface brightness dwarf galaxies} (LSBD) show the same behavior, but
require an even higher relative amount of dark matter; this is in contradiction to
cosmological DM models \cite{Blo:01, Bos:02}, as it is adressed in detail in \cite{Tas:02}.
The same holds for tidal dwarf galaxies \cite{Bou:07, Gen:07}.

\paragraph{Globular cluster distribution.} Rotation curves could be explained with
dark matter located in the disc, but there is clear evidence that the gravitational
potential obeys radial symmetry. The spatial distribution  and velocities of globular
clusters makes a dark matter concentration in the disk extremely unlikely, besides
other evidences like the magellanic stream \cite{Agu:01a}.

\paragraph{LMC and SMC.} Recently, the HST provided 3-D velocity measurements of
these objects known as satellite galaxies \cite{Kal:06, Bes:07}. The abnormal high
 value suggests either that we are wittnessing a coincidental 
fly by of the Magellanic Clouds at the milky way, or a conflict with the mass estimates
hitherto existing.

Thought the effectic distance is at the planetary scale, recently extracted data on 
eclipsing binary systems in the LMC could be useful for gravity tests \cite{Der:07}.

\paragraph{Local group, M31.}
The Milky Way and Andromeda (M31), the biggest galaxies in the local group,
are approaching each other much faster than can be explained by
gravitational attraction of the visible mass. The extra (dark) matter required
exceeds the visible one by a factor of about 70 \cite{Schn:06}. In this context, the recent
discovery of very distant halo stars in Andromeda \cite{Sar:06} is surprising, too.
Velocity dispersions of the local group are analyzed in \cite{Ekh:01}.

\subsubsection{Galaxy cluster scale}

\paragraph{The peculiar velocities} of galaxies in galaxy clusters were the first hint that
observed and expected (from luminosity and Newton's law) velocities did not match.
This was discovered as early as 1933 by Zwicky  \cite{Zwi:33} and called the
`missing mass problem'. Recent measurements are \cite{Sch:06}.

\paragraph{Gravitational lensing} confirms this result, since the observed light deflection
is much greater that the amount that could be explained by visible matter.
There is however a discrepancy in mass determinations for scales smaller than
300 $kpc$ (\cite{Schn:06}, p.~264).

\paragraph{Hot gas} in galaxy clusters is a further, independent confirmation
of the `dark matter' phenomenon. X-ray emission allows to
determine the temperature of intergalactic gas. Assuming that hot
gas being bound to the galaxy cluster (otherwise it should have
left the cluster), much more mass than the visible amount is
needed to explain the gravitational force that keeps it. A discrepancy
with $\Lambda$CDM predictions was claimed by \cite{Bro:08}.

\paragraph{Galaxy distribution} is an interesting testbed for gravity theories.
Excellent data are now provided by the Sloan Digital Sky Survey (SDSS) 
and the 2dF redshift survey. Both contain  positions and redshifts of
 galaxies, thus
yielding information on the 3-D structure. It is known for more than 20 years
that the distribution of galaxies is not really homogeneous.
Rather there seems to be a hierarchy of galaxy
groups, clusters and superclusters that concentrate on
twodimensional structures, while there are
large voids in between (`sponge structure') \cite{Gel:86, Gel:89}. 
Only for scales larger than 100 $Mpc$, 
the universe appears homogeneous (but see also \cite{Bro:90}).
In particular the presence of large voids has been puzzling to cosmologists \cite{Pee:01}
since they appear hard to explain by th $\Lambda$ CDM model \cite{Pee:07}.
There is an ongoing discussion whether the large scale structure
becomes homogeneous or possesses a fractal structure \cite{Syl:08}.

It is remarkable that the relation dark to visible matter is even higher for galaxy clusters
than for spiral galaxies. It seems that the larger the structures, the more dark matter is
needed to explain their stability.

\subsubsection{Cosmological scale.}

\paragraph{High-redshift supernovae.}

The determination of today's value of the Hubble constant $H_0$ is
a long story. In the 1990's, there was still a large discrepancy
between groups favoring a value of $80-100 \ km s^{-1} Mpc^{-1}$
based on several methods and the value of about $50 \ km s^{-1}
Mpc^{-1}$ which arose from quite accurate measurements of the
luminosity of supernovae of type Ia (the actual accepted value
is  $73 \pm 8 \ km s^{-1} Mpc^{-1}$ \cite{Fre:00, Rie:05}).\fo{It is a fact that
`generally accepted' may change quickly in cosmology. \cite{San:06} reports
$62 \pm 1.3 \ km s^{-1} Mpc^{-1}$, see the review \cite{jac:07}. John Huchra
maintains a list of $H_0$ determinations: 
$http://cfa-www.harvard.edu/~huchra/hubble.plot.dat$.} Then
\cite{Sup, Sup4, Sup5} and \cite{Sup2} independently announced
that the relatively too faint high-redshift supernovae should be
interpreted as an accelerated expansion of the universe. This is commonly explained
by postulating a new form of matter called `dark energy' (DE) that
acts repulsively.
It should be kept in mind that considerable data reduction has
taken place. There are hints that the chemical distribution of
elements causes brightness differences in SN explosions that would
affect systematically the SN Ia data; simulations are currently
carried out \cite{Roe:06}. While the data clearly exclude a
universe of ordinary matter, the question as to what `dark energy'
consists of is completely open. Laboratory tests could not give 
evidence for it yet \cite{Kap:07} and there are still discrepant
observations from X-ray data \cite{Vau:03, Lun:04}. 
A detailed look at the data
\cite{Rie:04} however shows that some surprising interpretation is
not yet excluded; the data is even compatible with an 
`empty'$\Om=0$\footnote{See \cite{Kow:08} for recent constraints.} universe.

\paragraph{Weak gravitational lensing} \cite{Btl:01} is a powerful tool to estimate mass
distributions on cosmological scales. These data 
indeed can constrain the long-range properties of gravity. 
The additional claim \cite{Whi} that alternative theories are ruled out by the data 
refers however to a special class of models which show 
different radial dependencies. Alternative approaches in general, not even
the concrete MOND proposal, cannot be tested, see \cite{Whi}, sec.~3. It 
seems that those results attack models with fixed-$r$ crossovers, something which
is already dead \cite{Agu:01a}.

\paragraph{Big Bang Nucleosynthesis (BBN).}
The big bang scenario with a hot, fast expanding universe allows
to make testable predictions regarding the abundances of light
elements that must have formed in the first minutes, in particular
helium and deuterium. Though doing calculations in this extreme
state of matter requires considerable extrapolation of physical
laws, the observed abundances in primordial clouds justify this
interplay of thermodynamics and nuclear physics. Since
gravitation should slow down the expansion, a tiny effect on
element abundances could be predicted due to gravity, too.
\cite{Cop:03} have used this to constrain the value of $G$, and by
assuming some temporal dependency, also to put limits on $\dot
G/G$ and on scalar-tensor theories of gravity \cite{Coc:06}.
To evaluate the significance of such a test for gravitational
physics, some remarks must be given. There is no doubt that the
amount of $He_4$ found in  primordial clouds backs the big bang
model in general. For further predictions, deuterium abundances
from QSO spectra are measured which are less accurate. It is then
assumed that the amount of $D$ did not change since primordial
times. Moreover, a lot of recent knowledge of particle physics,
such as neutrino oscillations enter the calculations. Reading
\cite{Kne:04} is useful to get an impression of the complexity of
what experts call the simplest case. All assumptions entering seem
plausible at the moment, but there are many of them. Additionally,
to interpret 
the missing baryon density, dark matter is needed.
Most importantly perhaps, the application of GR to this very
early, radiation-dominated phase of the universe, requires the
strong equivalence principle, which is not really tested
elsewhere.

\paragraph{Cosmic microwave Background (CMB).}
The COBE and WMAP missions provided unique data on the blackbody character \cite{Ben:03},
on the fluctuations and on the polarization of the CMB. Its very existence
is a very strong
evidence for the hot big bang scenario. From its power spectrum,
a couple of physical parameters can be estimated, such as the baryon-to-photon
ratio used in the BBN calculations \cite{Coc:03}, and imprints of modified gravity could
be found there \cite{Ser:06}. A critical point of view is outlined by \cite{Sha:05}.
 The most important result is that at the largest
scales, the  universe appears to be flat, or, in old-fashioned terms, kinetic and
potential energy are perfectly balanced.
An interesting aspect is that CMB  - for the first time
in the history of physics- introduced a physically meaningful
a preferred system. Of course, there is no conflict with Lorentz invariance.

The density fluctuations of the CMB are by far insufficient to explain
galaxy formation.  Large amounts of `dark matter' and additional assumptions
on its fluctuations are needed to match the observed galaxy distribution.

\subsection{Tests ordered by mass}
I will refer sometimes to the previous section, but it is useful to
have a look on gravity tests under different aspects.
If Newton's law with its symmetry is not assumed {\em a priori\/},
one first has to distinguish test and field masses. The independence
of test masses is called the (weak) equivalence principle (EP), but it
usually refers to a material independence. It is hard to imagine
a scale dependence without being in conflict with logic, but
a strict test is possible only if the mass of the test particle is
precisely known.

\subsubsection{General remarks and small scales.}

\paragraph{Linearity and superposition.}
More interesting seems to wonder about a dependency of the law of
gravitation on the amount of the field mass. Of course, in the case
$m=M$ the law should become symmetric again, but a deviation for
large masses would not necessarily violate the EP, at least
within its experimental constraints which mostly apply to test masses.
It is not that common to wonder about such a dependence because
it is completely out of our theoretical expectations. For instance,
mass distributions with spherical symmetry could not automatically
be replaced by point masses
any more. Moreover, the method of mass integration as such
bears a linearity that would be put into question
by a mass dependence. We are used to the nonlinearities arising in
GR,
and automatically infer 
that gravity must be linear in the Newtonian limit, that is, the superposition
principle holds. The claim that any physical theory must be linear in the
weak-field-limit reflects some of our experience but cannot be rigorously
proven. Strictly speaking, we perform an extrapolation of our simple
mathematical methods which must be tested.
While observations regarding  distance test the exponent $2$ in (\ref{newton}),
it turns out that the exponent $1$ on $M$ is much more difficult to prove.

\paragraph{Light.}
Gravity obviously acts on particles without rest mass, too. Inserting
the photon energy $E= h f$ and $ m= E/c^2$ however would yield only
the wrong (half) value of a deflection due to gravity, thus light
deflection needs to be understood relativistically. An effect of photons as field
masses would be postulated by the strong EP. There is no test available.

\paragraph{Atomic scale.}
Though there is hope to measure gravity for neutrons as test particles
by interferometry \cite{BeL:06}, the effect of an elementary particle or atom
as field mass seems unconquerable
small.

\subsubsection{Laboratory scale.}

\paragraph{Torsion balance experiments.}
One has to skip 25 orders of magnitude in both field and test masses
to get the first measurable effect of gravity. Torsion balance experiments
usually work with equal masses in the range of 10 kg. Modern versions
\cite{Gun:00,Qui:01} provide the most accurate measurements of $G$.

\paragraph{Heavier field masses.}
One SG experiment \cite{Bal:05} used $280 \ kg$, the free fall method
\cite{Sch:99} $500 \ kg$, and \cite{Sch:02} about $13000 \ kg$
as field masses, and the lake experiments \cite{Bal:01, Hub:95}
reach $10^{7} kg$. The accuracy for determining $G$ however usually
decreases while using heavier field masses.

\subsubsection{Geophysical scale}
The measurements with moving gravimeters considered above must be
seen as tests with an effective mass of a part of the upper crust
surrounding the gravimeter. A crude estimate of the volume
corresponding o the greenland ice experiment \cite{Zum:91} with a
depth of $1673 \ m$ would be some $km^3$, reaching a mass scale of
about $10^{12} kg$. Though there were claims for deviations from
Newton's law in this regime, the experiments remain difficult to
interpret.  Usually, the gravitational mass determination enters
all geophysical models of the core and mantle composition. It is
however interesting to try a purely geophysical mass estimate by
the earths volume and the suspected chemical distribution
\cite{Mik:77}. Important information about that can be obtained
from  seismic waves. According to the common model, there is a
density jump from $6-10 \ g \ cm^{-3}$ at the core-mantle boundary
(\cite{BeS7}, p.~30).

\subsubsection{Solar System scale}

There is quite a big gap between the previous tests and those arising from orbital data
of celestial bodies. Strictly speaking, a possible field mass dependence of Newton's law
(i.e. an exponent $M^{\a}$ with $\a \neq 1$ in eqn.~\ref{newton}) 
 can hardly be detected by satellite trajectories, since the same data are used to measure the
mass. Independent assumptions on the densities of the sun and its planets
usually lead to crude estimates only.
While the Keplerian constants from different satellites (planets) are very accurate
tests of the inverse-square-law in distance, they cannot reveal an $\a$
slightly different from $1$. Thus, as far as field masses are concerned,
in the range from $10^{23} kg$ (moon) to
$10^{30} kg$ (sun) no test with significant accuracy exists.

\subsubsection{Galactic scale}
As a matter of principle, the same problem of an independent mass
estimate holds for galaxies, too. The difference is that even very
crude estimates like the assumption of a solar mass-to-light ratio
by far do not match the dynamically determined mass which is in
the range of $10^{40}-10^{44} \ kg$. This phenomenon could be
explained either by forms of matter which interact gravitationally
only (`dark matter') or by failure of Newton's law. Contrarily to
opposite claims, a mass dependence $M^{\a}$ with $\a<1$ does not
imply a violation of the equivalence principle, since the latter
holds for test particles. From galactic rotation curves, a
dependence $M^{\frac{1}{2}}$ has been suspected \cite{Agu:03}.

\paragraph{Radial and tangential velocities.} The data of solar system observations consists of
precise orbital data in three dimensions. The situation is very
different on larger scales. While radial velocity measurements can
easily performed by Doppler methods on galactic and cosmological
distances, measuring velocities perpendicular to the line of sight
was almost impossible for a long time. This changed very recently
for the galactic scale when VLBI and HST detected secular shifts
of galactic objects on the microarcsecond 
level. Many of these results, however, were surprising in the
sense that they did not match the predictions of Keplerian
ellipses \cite{Kal:06, Bes:07, Ruz:08}. The problem at the galaxy and galaxy cluster scale is
that our knowledge comes from a snapshot of tens of years and we do
not really know about the secular dynamics. The common picture of
stable and virialized 
systems is an extrapolation of accepted
theories of gravity. Observations on possible radial flows in
galaxies are discussed in \cite{Won:04}.

\subsubsection{Cosmological scale}

For a long time, mass  (or equivalently, density) estimates were
used to decide the question whether the universe will stop its
expansion due to gravitational attraction and recontract (closed
universe, $\Om>1$) or expand forever (open, $\Om<1$ ).
Interestingly, many measurements suggested a state just in between
($\Om=1$). The recent WMAP data also indicate the puzzling case
$\Om=1$. The problem is that this nice picture is completely
screwed up by the SNIa data that show an {\em accelerated\/} expansion 
of the universe
(while before the debate was going on how much it was
decelerated). Though since Newton attraction is quite a
characteristic property of matter, many physicists do not have
problems to assume a repulsive interaction to DE, while others
admit that `dark energy' might be just a name for something 
we do not understand.

\subsection{Tests ordered by acceleration}
This seems uncommon but it may be useful to consider the different orders of magnitude in
which the respective tests are performed. We should be aware of an
inadequate extrapolation of physical theories.

\subsubsection{Strong and intermediate scale}

\paragraph{Black holes.} Though GR predicts the existence of black holes (BHs),
the current, mostly indirect observational evidence for BHs is not a quantitative test of GR.
Spectacular measurements of stellar orbits around the center of the milky way 
\cite{Sch:02} allowed to deduce a mass concentration that indicates a
supermassive black hole (SMBH). 
This is however not a quantitative measurement of the Schwarzschid radius
$r_s=\frac{2 G M}{c^2}$ independent from the validity of GR, since $r_s$
in this case would be still smaller than the perigee distance by a factor of about
1500. This factor may be decreased by VLBI observations \cite{Doe:08}.
The transition from neutron stars to BH's is dscussed in \cite{Fre:08}.
A review of strong field tests is \cite{Psa:08}.

\paragraph{Neutron stars.}
At the surface of neutron stars, accelerations up to
$10^{12} \ m s^{-2}$ can be reached. Compared to the acceleration
of an electron a the Bohr radius, $10^{23} \ m s^{-2}$, this is
not really high, but of course, no gravity test can be performed.
The highest possible (orbital) accelerations in binary star systems like
the famous $PSR \ 1913+16$ are in the order of $50$ - $350 \ m s^{-2}$,
the double Pulsar $PSR \ J0737-3039$ \cite{Ior:06g} reaches $100 \ m s^{-2}$.
Indeed, the periastron advance of $PSR \ 1913+16$ is a qualitative proof
that the well-known Mercury 
 perihelion advance exists in the strong acceleration
regime, too. Assumed that GR is correct, this allows a unique mass determination
of the binary system. Moreover, the decrease of the rotation of the system
is in excellent agreement with the energy loss predicted by the radiation
of gravitational waves. Reviews on testing gravity with binary and milisecond pulsars are
\cite{Dam:07, Lor:08}.

\paragraph{Planetary surfaces.}
While typical accelerations are in the order of $10 \ m s^{-2}$,
local gravity measures on earth with superconducting gravimeters have reached
an accuracy of $10^{-12} \ m s^{-2}$. Due to a variety of geophysical effects,
it is difficult to perform gravity tests. All absolute $G$ measurements have been conducted
under these `background' conditions. However, the experiments \cite{Gun:00, Qui:01}
and \cite{Sch:02} measured the gravitational accelerations in perpendicular directions.

\paragraph{Satellite orbits.}

Satellites like LAGEOS act at an orbital acceleration of about
$2.5 \ m s^{-2}$, therefore this regime is tested extremely well.
The flyby anomaly, however, occurs at similar experimental
situations, though on hyperbolic orbits. The amount of the
corresponding anomalous acceleration is not measured very well
yet.

\paragraph{Solar system scale.}
The orbital accelerations of the planets range from $3.9 \times
10^{-2}$ to $6.6 \times 10^{-6} m s^{-2}$ (Neptune). Newtonian
gravity is tested extremely well in this regime, and, as far as
planetary orbits are concerned, no deviations are found above
$10^{-10} m s^{-2}$ \cite{Ior:06c}. This is important since the
Pioneer anomalous acceleration  $a_p= 8.74 \times 10^{-10} m s^{-2}$
was measured for a comparatively light spacecraft on a hyperbolic
orbit.

\subsubsection{Weak (galactic) scale}

\paragraph{Galaxy rotation curves and MOND.}
This scale is the most interesting one and the reason for
analyzing gravity under the aspect of acceleration strength. The
most recent observational overviews are \cite{Bos:03, Bos:03a, Sal:07a}. 
A detailed look however shows that 
many observational facts can hardly be explained by {\em any\/}
`dark matter' theory \cite{Sel:00}. Further overviews focussing on that problems are
\cite{Eva:01, Tsa:02}.
As an alternative, galactic rotation curves
gave rise to speculations on a modification of Newton's law.
However, all proposals that tried to modify it with respect to
distance, ran into a tremendous mismatch with the data, which is
outlined in detail by \cite{Agu:01a}. The very unusual proposal of
Modified Newtonian Dynamics (MOND) \cite{Mil:84, Mil:98}, received
with scepticism, was based on the remarkable observation that the
assumed failure of Newton's law occurred at a fixed {\em
dynamical\/} scale in the order of $10^{-10} m s^{-2}$. The
concrete proposal of an effective acceleration 
\be g= \sqrt{a_0 \frac{G M}{r^2}} \label{MOND}
\ee 
with the fitted parameter $a_0 \approx 1.1 \times 10^{-10} m
s^{-2}$ indeed matches most of the galactic rotation curves with a
reasonable accuracy \cite{Mil:84,Mil:98}. As in the case of the
Pioneer anomaly, the approximate coincidence of $a_0$
with $c H_0$ (you may divide by $2 \pi$)  attracted attention. There 
are even phenomenological models encompassing both effects \cite{min:06}.
Laboratory \cite{Ign:07} and solar system \cite{Ior:07a} tests for MOND 
have been proposed, while
artificial planetary systems in space may reach this small acceleration
scale \cite{Sha:06}.
Of course, MOND is unusual and it is quite boring to itemize the
fundamental principles of theoretical physics it contradicts
\cite{Sco}, and there are observations where MOND has problems, too
\cite{Agu:01}. So what~? I don't know if anybody thinks that MOND
is the last word in gravitational physics. Its indisputable merit
however is to have attracted attention to the fact that Newton's
law is poorly tested for accelerations below $10^{-10} m s^{-2}$.
The approximate agreement of $a_0$ with $c H_0$ is either a
coincidence invented by nature to fool astronomers or a proof that
we do not understand gravity yet.

\paragraph{Globular clusters.}

Globular clusters, though orders of magnitude smaller than spiral
galaxies,  have comparable accelerations at their boundaries. The
recent observations \cite{Sca:06a, Sca:07} that flat velocity dispersion profiles occur in
globulars, too, have a great impact. It is not important whether this
is properly described by MOND (though $a_0$ is the same), but if
the results are confirmed, they will bring the dark matter model into serious
trouble.

\paragraph{Galaxy clusters.}

Obviously the accelerations in galaxy clusters are usually even weaker than the
centripetal ones of rotation curves. Again, the gravitational accelerations necessary
to keep clusters together would be much greater than those suggested by visible
matter. The data do not allow to distinguish between the dark matter hypothesis
and a failure of Newtonian gravity (with instructive comments, \cite{Sel:00, Sel:04}).

\subsubsection{Cosmological scale.}

Though there is no direct measurement of the large-scale
accelerations occurring during cosmic evolution, the predictions of
 Friedmann-Lemaitre cosmology can be tested. As outlined above,
there are some contradicting results with no satisfactory solution
at hand yet. To get an order of magnitude, one may assume the
masses to be accelerated to the Hubble velocity during $H_0^{-1}
=13.4 \ Gyr$, ending up
with $c H_0 \approx 7 \times 10^{-10} m s^{-2}$, again in the weak-acceleration regime. 
If Newtonian gravity turns out to be wrong here, one should search
for a different explanation of the observational puzzles of
cosmic evolution, too.

\subsection{Collection of funny coincidences, problems and results under debate}

\bq
{\it `Everybody believes the experimenter - besides the experimenter himself'.}
\eq

\paragraph{The Tully-Fischer (TF) relation} is an empiric law \cite{TuF}
that relates the total luminosity
$L$ of a galaxy to the maximal rotational velocity observed by Doppler shifts
\be
L \sim v^{\b},
\ee
whereby $\b$ ranges from 3.5 to 4.5. The TF relation is quite accurate, thus it has been
used to determine the Hubble constant, too. A theoretical reason for its origin does not
exist in standard cosmology. A similar relation has recently been observed for
the radial velocity \cite{Yeg:06}.

\paragraph{Further relations.} The $L_{bulge} -\s$ or Faber-Jackson-relation \cite{Fab:76} 
relates the luminosity in the central region to the velocity dispersion $\s$ of
the respective stars. Even more surprisingly, there is a correlation of the central black 
hole mass $M_{BH}$ of a galaxy and the velocity dispersion, the 
$M_{BH}$-$\s$ relation \cite{Tre:02, MBH}.
While the details of these dependences are under discussion, the pure existence of relations 
between these three quantities are an unexpected regularity which miss a theoretical explanation.


\paragraph{Density waves.} It has long been noted that the visible spirals of
galaxies cannot rotate rigidly, because differential rotation would swipe them out soon,
i.e. after a few hundred million years.
The phenomenon is commonly explained by acoustic waves that in the compressed
regions form bright, short-living stars that lead to a greater visibility of the spiral arms.
A direct proof of this theory has not been given yet (cfr.\cite{Sel:99}),
and one may ask how the galaxies %
maintain their ability to supply the star forming regions after
dozens of density wave passings. The formation of jets is not fully understood as 
well \cite{Lyn:02}.

\paragraph{Barred Galaxies} show a couple of results that are poorly understood yet.
These discrepancies and others are discussed in \cite{Eva:01}. An extensive review
on barred galaxies is given in \cite{Sel:93}.

\paragraph{Numerical simulations} of galactic structure formation predict thousands of
dwarf galaxies in the vicinity of large spirals like the milky way. There are only 
35 objects of that type observed currently.

\paragraph{The quadrupole-anomaly} has been detected in the COBE and WMAP 
data of the CMB \cite{Teg:03, Lam:06}.
It means that the fluctuation amplitude is much lower than
expected, and moreover, the anomalous small quadrupole and octupole component
are aligned to the ecliptic. 
The alignment to the ecliptic, rather than to the galactic plane,
 makes it unlikely that an artefact appeared due to the need of 
removing foreground signals from the data.

\paragraph{An increase of the astronomic unit} of about $10 m$ per century
has been reported recently \cite{Kra:04, Pit:04, Sta:04, Lam:06}. If this result
which is still under debate is confirmed,
it will be quite hard to find a conventional mechanism that explains it.

\paragraph{The third-parameter problem} means the impossibility to account for
the observed variability in HRD diagrams of globular clusters by means of  two
parameters. `Either there is some fundamental failing in our
understanding of stellar evolution, or there must be some other
factor beside the metallicity 
and age which dictates the
properties of globular clusters' (\cite{BinGA}, p.~352).

\paragraph{Absorption lines of quasars} have been used to estimate the fine structure
constant $\a$ in the early universe. Webb \cite{Web:98, Tza:06} deduced a different value for $\a$, but
a corresponding rate of change would be to large for being undetected by actual precision
measurements \cite{Fis:04}, see also \cite{Ste:08}. 
The debate is going on, in particular which other constants of nature can be
affected by a change, if any \cite{Uza}.

\paragraph{Quasar statistics} reveal that they appear more frequently near the sight line
of foreground galaxies, at least this is claimed by
Arp, a highly skilled cosmologist \cite{Arp:97,Bur:05}. The cosmological 
model favored by him seems too exotic to be credible, but
in his nevertheless recommendable book \cite{Arp} 
he soberly observes lots of facts that do not fit into the commonly
accepted picture.

\paragraph{The astrometry satellite HIPPARCOS} provided data of unique 
precision\footnote{This precision will still be increased by the GAIA mission.},
and the parallax method of determining distances is considered the most direct and
best for small distances. However, the parallactic distances were considerably smaller
than those measured by other  methods. Recently, \cite{Zwa:04} presented a distance
measurement using spectroscopically gained orbital data of a binary star system, which
can be considered similarly direct (`geometric') than parallaxes. The discrepancy
with the HIPPARCOS distance awaits for an explanation. Problems with HIPPARCOS
distances are also discussed in \cite{Krs:00}, see also \cite{Bru:08}.

\paragraph{Galaxy surfaces.} The `surface' of the Milky way can  be approximated by
$2 r_g^2 \pi$ with $r_g= 10000 \ pc$. Assuming as a crude estimate $10^{11}$ Galaxies of this
kind in the universe leads to a total `surface' in the range of $10^{52}-10^{53} \ m$ which
coincides with the surface of a sphere with $r = c H_0^{-1}$. Given that galaxies are assumed to be
unchanged in shape (`virialized') for a long time,
 this is an astonishing property of the present epoch. 

\paragraph{The `direct detection'} of dark matter was a recent claim
\cite{Clo:06}. These interesting  observations however do rather raise new questions
than resolve the DM riddle \cite{Ang:07, Bro:08}.

\paragraph{Gravitational waves} have not been detected yet, notwithstanding
enormous efforts, see \cite{Row:00} for a review. 
The evidence provided by the slowdown of the pulsar $PSR \ 1913+16$ 
for this prediction of GR thus remains indirect (cfr. \cite{Dam:07, Lor:08}).

\paragraph{Cryogenic sapphire oscillators} show drift rates linear and in the same 
direction for many years \cite{Tob:06}. Since the order of magnitude is $10^{-13}$ to $10^{-14}$
per day, one could speculate on a possible connection to the Hubble time (see also \cite{Biz:04}).

\paragraph{Gravity shielding.}
There are quite suspicious reports on gravity shielding over superconducting coils
\cite{Pod:97,TPSM}. This seems extremely unlikely; one the other hand there is no reason why
one should not try to repeat a quite simple experiment.

\subsection{Overview}
\bq
{\it `It is the stars, the stars above us, govern our conditions' (King Lear, Act IV, Scene 3)}
\eq
\begin{figure}[h]
\includegraphics[width=14cm]{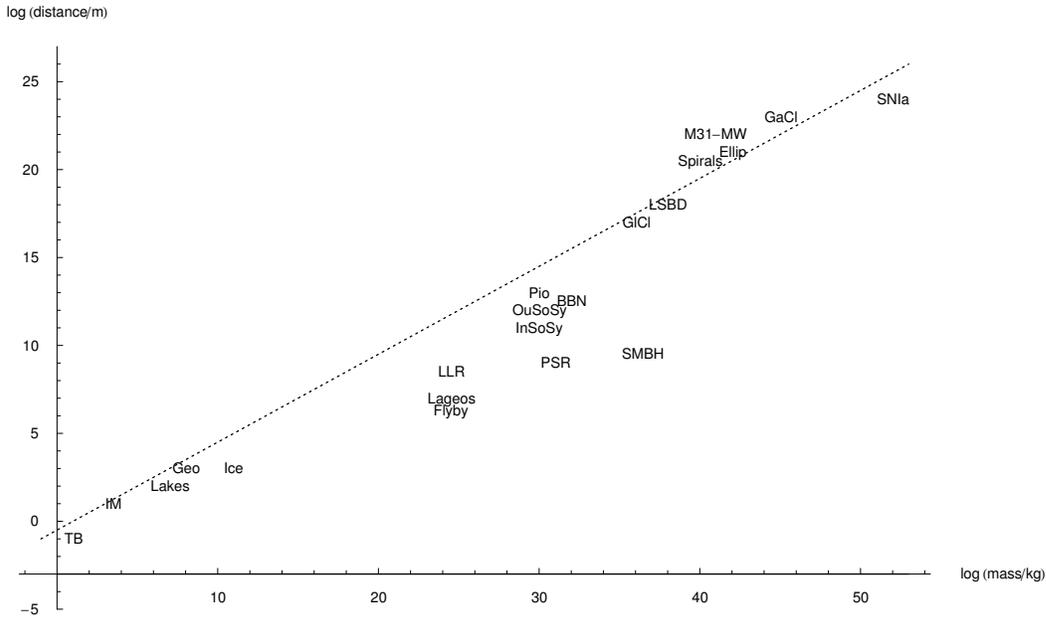}
\caption{Overview on data domains,
length is upwards, masses to the right. BBN is shown at the
then-horizon and the corresponding  density. For the SMBH, $10^6 \ M_{Sun}$ is
assumed, but there is no test. The logarithmic units may yield to an 
underestimation of the degree of extrapolation.
Actually there are two big gaps. In the right upper region, no gravity tests at all are
available. The dotted line corresponds to the acceleration
$c H_0 \approx 7 \times 10^{-10} m^2 s^{-1}$. Below that regime (above the line)
there are no significant tests of gravity.}
\label{grmap}
\end{figure}

\paragraph{The gaps.}
Fig.~\ref{grmap} gives an overview of the orders of magnitude involved by
tests of gravity. Roughly, there are three groups: left (below), middle and
(upper) right. The left group involves tests on small scales, and only here
absolute $G$ measurements are possible. Then a considerable gap follows,
and one should not forget that as far as masses are concerned, anything else is extrapolation.
The middle group is the well-tested
regime of celestial mechanics, on which our belief in Newtonian gravity and
GR relies, but strictly speaking it tests Kepler's law rather than Newton's.
The Pioneer and flyby results are not explained yet, however.
The upper right group represents gravity tests from the galactic to the
cosmologic scale. None of them yields undoubtable
evidence for the validity of Newton's law, i.e. i must be backed by additional
assumptions like DM or DE. In summary, \\
(1) Tests of the field mass dependence are entirely determined by only 1-2 independent types
of experiments on small scales. Given the relatively poor accuracy, the extrapolation
to the following is courageous. \\
(2) The data in the middle group is well-tested, but keeping in mind the logarithmic diagram,
further extrapolation is not backed by the observations, not even for Kepler's law.


\paragraph{A general failure of gravity at low accelerations~?}

\bq
{\it The fault, dear Brutus, is not in our stars, but in ourselves' (Julius Caesar, Act I, Scene 2)}
\eq

In the last section a variety of anomalies was discussed. The most puzzling fact is that
accelerations of the order $10^{-10} m s^{-2}$ occur in several completely different circumstances.
If we try to give an order of strength of evidence, this involves
(1) Galactic rotation curves,
(2) The Pioneer anomaly,
(3) Globular cluster data.
Moreover, all the other  unexpected phenomena consistently occur 
below that acceleration regime. The enigmatic coincidence with $c H_0$ 
is an additional strong hint that the discrepancies have a common origin
and may lead to a general failure of Newton's law in the weak-acceleration
regime.

\section{The agreement with theory}

\bq
{\it `The great tragedy of Science--the slaying of a beautiful hypothesis by an ugly fact.'
- Thomas H. Huxley}
 \eq

This is an ironic version of Popper's criterion of falsifiability, 
indeed a good attribute for testing whether a theory is scientific or not. For the
development of science however it tells only how things {\em should\/} work.
In practice, established theories which frequently are considered `beautiful',
do not die suddenly when confronted with ugly facts, they rather affiliate them
by becoming extended `models'. 
This process was described by the philosopher Kuhn \cite{Kuh:62}
as `normal science', which tries to understand the observations within an accepted framework.
When the anomalies pile up to an obviously too complicated scheme, a paradigm shift
takes place that may lead to a `scientific revolution'. What is `obvious' 
and whether the crisis has begun or not is not easy to decide. The last occurrence in history, 
the Copernican paradigm shift, tells us in retrospect only that scientists may be blind
for a long time.

\subsection{Standard cosmology, $\Lambda$CDM}

\bq
{\it `Cosmologists are often in error, but never in doubt' - Lev D. Landau.}
\eq

\paragraph{Friedmann-Lemaitre cosmology and `dark energy'.}
After having discovered the relation between gravity and geometry,
Einstein wasn't aware that his field equations did not permit the
static solutions he desired. This was shown by Friedmann, and
Lemaitre started to develop the model which is nowadays called
big-bang-cosmology. To justify his favored steady-state-model,
Einstein introduced the cosmological constant $\Lambda$, but after
having realized that Hubble's data supported an expansion of the
universe, he later called $\Lambda$ the `biggest blunder' of his
life. The high-redshift supernovae, indicating an accelerated
expansion, led to a renaissance of $\Lambda$, now called `dark
energy'. To claim Einstein was right and praise his ingenuity
however is in this case misleading - apart from the fact that a
couple of his interesting ideas still are considered blunders by
modern-minded people. Einstein was ready to introduce  a
mathematical complication in order to save a physically simple
model, because of his deep-rooted conviction that laws of nature
must be simple in a physical sense. Today's dark energy instead is
just an additional free parameter used to adjust a more and more
complicated observational situation \cite{Tur:02, Krs:04}. I think that Einstein had
preferred to admit a blunder rather than being a   chief witness
for $\Lambda$.

\paragraph{Dark matter.}
\bq
{\it `There is certainly not a lack of chutzpa in extragalactic astronomy' - Margret Geller} 
\eq

There is
overwhelming evidence that the `dark matter' {\em phenomenon\/} exists on
various scales. 
While the DM hypothesis explains
galactic rotation curves at a first glance, maintaining
it in the light of more recent results seems to rely on the absence
of detailed knowledge. 
Moreover,
the absence of any decline suggests that the dark halos continue to regions
not accessible by current technology. Cosmologists should be prepared that
the `fraction of DM' turns out to be just a measure of telescope resolution.
Then, it is strange enough that
the relative amount of dark matter seems to increase with the size
of the structures (galaxy clusters contain much more than
galaxies), but dark matter can clearly not explain the recent anomaly
found in globular clusters or even the Pioneer anomaly. I am
curious to future  priority claims for introducing the terms `dark
molasse' and `dark substance', which, with an appropriately chosen
distribution, could describe both
phenomena satisfactory. 
Details of what has been proposed as candidates
for DM cannot be addressed here. Massive compact halo objects (MACHOS), such as
brown dwarfs, would have been detected by microlensing if they
existed in the required amount. While the question of hot (fast,
relativistic particles) or cold dark matter seems to be answered
in favor of the latter, no reasonable candidate particle has been
found in the laboratories yet. To some people, postulating
neutralinos (or elsewhere, photinos and axinos) still gives hope
for a possible explanation, since Italian language will hardly run
out of diminutives. Following this road, poorly understood new 
results \cite{DAMA:08, Tom:08} can be readily interpreted as evidence for 
various dark matter particles. On the other hand, this development reminds 
strongly from Ptolemaic epicycles \cite{LoC:08}.

Dark matter  however remains at least an urgent need in the codes of 
programs simulating cosmic structure formation. The
fluctuations in the CMB are by far insufficient to explain the clumping that
corresponds to the concentration of galaxies. Dark matter instead, numerically much cheaper than ordinary
matter, clumps so easily as long as we do know about its properties. Isn't that a nice proof of
its existence~? Or should we admit that we do not understand structure formation~?

It is time to put into question the obvious idea of explaining deviations 
from Newton's law with gravitationally interacting matter not yet detected. From the
point of view of scientific methodology, the last success of a
dark matter theory was the discovery of Neptune. 

\bq
{\it `Once the error is based like a foundation stone in the ground, 
everything is built thereupon, nevermore it returns to light.' - Rudolf Clausius}
\eq

\paragraph{The Flatness problem.}
Measurements of the density parameter $\Om$, in particular from the recent
WMAP data, yield a value very close to 1. However, evolution models
of the universe would predict a strong drift of this value, once there is a minute
deviation from 1. Therefore, at early times, $\Om$ must have been as close as $10^{-60}$
to 1, and the question arises how this fine-tuning
of a measuring value occurred, or if there is a theoretical reason
beyond.

\paragraph{The Horizon problem.}
The finite speed of light implies that we are not able
to see parts of the universe further away than the 
distance light could travel since the big bang. This
implies that, at the time being emitted, CMB photons at different
positions (separated by more than $1 \deg$) in the sky could
not `know from each other', i.e. there was no causal contact.
The question arises how a common mechanism suggested by the highly uniform
temperature, could occur.

\bq
{ \it `Inflation is a fashion the high-energy physicists have visited on the cosmologists; 

even aardvarks think that their offspring are beautiful.' - Roger Penrose}
\eq

\paragraph{Inflation}
is called that the hypothesis that the universe expanded by a factor
$10^{40}$ at the period around $10^{-34}s$ after the big bang. Indeed, it provides
an explanation for the horizon and flatness problem. Looking at the
uniformity of structures, such arguments appear superficial (\cite{Pen}, ch.~28.5):
\bq
`whatever this generic singular structure is, it is not something we can expect
to become ironed out simply because of a physics that allows inflationary processes.'
\eq
While inflation becomes more
and more a component of the standard model, many physicists feel uncomfortable
with it. The problem is not - as it may seem - that there is no physical mechanism
for such a superluminal process, actually there are many of them -
but no one will ever be tested.
How to observe $10^{-34}s$ after\fo{The only reasonable
measurement of time that can be performed in physics is based on the frequencies
of atomic or nuclear transitions. Since this could be at best in the range of
$10^{-15}s$, such theories will never come down to earth.}
 the big bang, which is still 380000 years
before the last scattering surface of the CMB \footnote{See \cite{Efs:07} on the vague 
possibilities of testing inflation with the CMB.} Inflation does not have a problem
with explaining phenomena, it has the problem that it explains almost everything.
The question
which observation would be incompatible with inflation, was raised by \cite{Bar:97}:
\bq
`This elasticity has diminished the faith of the general astronomical community
in inflation, and even led some  researchers to  question whether
inflationary cosmology is a branch of science at all'.
\eq
It is frequently claimed that inflation `predicted' a flat universe. However,
before the BOOMERANG and WMAP data ultimately measured the curvature $K \approx 0$,
numerous versions of inflation predicting $K \neq 0$ were circulating.
But most importantly, we cannot act as if $K=0$ were a measuring value like other numbers.
The observed flatness seems to be a very deep, cogent feature of the universe, 
for which an inherent theoretical reason should exist, not just
an ad-hoc, qualitative mechanism.
Of course, there could be some exotic topology in the WMAP data (still to be detected)
that would bring inflation into trouble, but it is hard to imagine that people were unable
to fix that by some modified or extended version... To conclude, the theory of inflation,
if not the origin of a never ending universe, is at least a starting point
for an inflation of theories never tested.

\paragraph{Not really an alternative theory but...} the flatness and the horizon
problem arises, because we believe in Friedmann-Lemaitre
(Newtonian) cosmology. At a constant expansion rate, more distant
objects have a greater relative `velocity' (Hubble's law). The
distance that corresponds to $c$ is called horizon. If the
expansion is slowed down by gravity, new objects drop into the
horizon. If there was no gravity acting, there is no slowdown, no
horizon increase and no problem. Instead of believing in a theory
relying on the constancy of $c$, then running into problems, and
resolving these problems by postulating an expansion $v \gg c$, isn't
it just more honest to say that standard cosmology is incompatible
with observation and we do not understand yet  cosmic
evolution~?

\begin{figure}[ht]
\includegraphics[width=10cm]{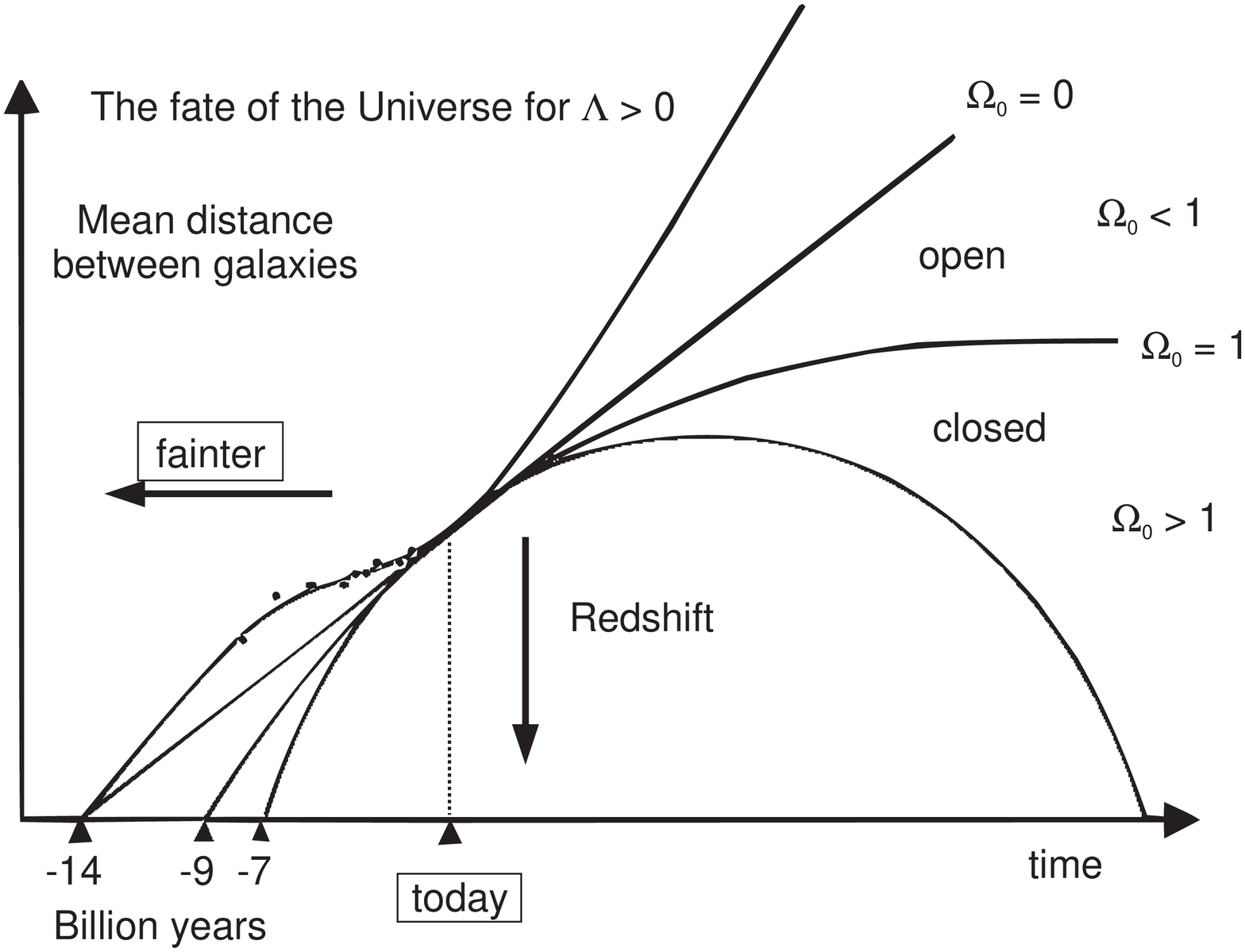}
\caption{The coincidence problem}
\label{coin}
\end{figure}

\paragraph{The coincidence problem} is best explained in a visual manner (see fig.~\ref{coin}).

While conventional models of matter (closed, $\Om=1$, and open)
seem to be excluded by the SNIa data (dots), the dark energy term
predicts an evolution of the universe corresponding to the above
line $\Lambda >0$. But isn't it strange that the `actual' tangent (Hubble's
constant, corresponding to $\Om=0$) hits the curved line
($\Lambda >0$) at the origin (big bang)~? Or in other words: we
seem to live in a very preferred period\fo{\cite{Krs:07} outlined this 
from a funny point of view.}, since the inverse current
expansion rate coincides with the time past since the big bang.
Let $p$ be the probability that the evolution of intelligent life
takes place in {\em such\/} a period, and $q$ the probability that
our actual understanding of cosmic evolution is quite complete,
it's up to the reader to decide whether $pq >(1-p)(1-q)$ holds.

\paragraph{The PPN formalism.}
While the standard model fits free parameters to data otherwise unexplainable,
there are many ideas of suggested violations of current theories which have not
been observed yet. A systematic method to  embrace
these constraints is the PPN formalism \cite{Wil:71, Wil:93, Wil:06},
which defines several parameters that could indicate certain kinds of violations.
This is a very useful tool for maintaining an overview on precision tests.
To take the values of the parameters in agreement with the standard model as support 
for its validity,  is however a circular argument.
Besides the fact that such a list of parameters can hardly be complete 
(e.g. \cite{Ior:07}), a nonzero violation parameter may easily turn 
into a new free parameter of an 
extended standard model. PPN, keeping the tradition  of fitting 
parameters is thus rather a seduction for future model makers than a helpful
indication that a fundamental new theory may be needed. Sometimes one can hear that
the PPN parameter space volume left by precision tests does not leave space
for extensions of gravity \cite{DeF:07,DeF:07a}.
It is well-known that the Ptolemaic model gave an accurate description of planetary orbits,
that means the `post-Ptolemaic-parameters' were close to zero. But it is
also well-known that the Copernican revolution did not take place in
a small `volume' of such a parameter space -  actually Newtonian gravity 
was something fundamentally deeper, not an extension of an existing model.
Taking the PPN formalism too serious could make us blind for such a possibility.
Further general comments on the standard model are given in \cite{Krs:04, Bay:06, Dur:08}.

\subsection{Alternatives}

\bq
{\it `Nobody believes the theorist - besides the theorist himself'. }
\eq

This is going to be the most subjective and incomplete section,
but it seems obvious to me that there is more need for a review of
observations than of theories.  GR was an extremely successful
extension of Newtonian gravity, but science requires the agreement
with observation, since `the danger of judging a theory on the
basis of elegance, simplicity or perfection is obvious' (R. Dicke
\cite{Dic:57}, p.~363). If a Copernican 
revolution is needed, it could well be that some Aristarchus had
already appeared, thus it might be smart to study some
of the old ideas. Publications on physical theories are
inevitably influenced  by fashion like other creations of
mankind; the extremely high quality of the present observational
data justifies indeed to neglect some older data. This led however
to an overestimation of current theoretical approaches with
respect to genius thinkers of the 19th and 20th century, a kind of
arrogance of the presence. 

At the same time, the standards what is
to be considered a good physical theory \cite{FeyQED} seem to be
in erosion. Einstein built a theory on the equivalence principle.
Nowadays we don't worry about deep principles, rarely there are
theoretical results, and many conferences entitled `Trends on ...'
take place.

\paragraph{Science and the Planck scale.}
It is common folklore
 that gravity needs to be unified with quantum mechanics,
and everybody, independent of loop or string preference, 
claims this must occur at the Planck level $l_p= \sqrt{\frac{G h}{c^3}} \approx 10^{-35} m$,
which protects comfortably 
from being falsified by experiment
\fo{It seems that in quantum gravity approaches this is slightly different. As 
Smolin \cite{Smo:06} points out, sophisticated proposals have been 
made how to approach the Planck level,
while string theories seem not to care any more about falsifiability.}.
The problem is that $l_p$ is deduced from $G$ and therefore 
significance of the argument relies on the correctness
of conventional theories of gravity. A theory that puts into practice Mach's principle,
could construct a relation of $G$ to the mass distribution of the universe and therefore
reveal the Planck level as an artifact. Many theories would lose then even a hypothetical 
contact to experiment.
For instance, another way to
look at the flatness problem is that potential energy equals kinetic energy, or,
simplified,
\be
\sum G \frac{m_i m_j}{r_{ij}} \approx \sum m_i v_i^2. \label{potkin}
\ee

Indeed,

\paragraph{Ernst Mach} suggested that the gravitational interaction had its origin in
the distribution of matter in the universe.\fo{See \cite{BaP} for an overview on Mach.}
  One possible idea to put Mach's principle into practice
is to modify (\ref{potkin}) and express $G$ as a function of the positions and velocities
of all masses in the universe. \cite{Sci:53, Lyn:95} and \cite{Bar:02} are
intelligent approaches in this direction.

\paragraph{A variable speed of light} was another interesting `blunder' of Einstein.
Those who think that such ideas contradict relativity or even worse, pure logic,
should remember that Einstein in 1907 wrote:

\bq
The constancy of the velocity of light can be maintained only insofar
as one
restricts ... to ... regions with constant gravitational potential...
\eq
and in 1911,

\bq `From the proposition which has just been proved, that the
velocity of light in the gravitational field is a function of the
place, we may easily infer, by means of Huyghens's
  principle,
that light-rays propagated across a gravitational field undergo
deflexion'. \eq

Recently, Ranada \cite{Ran:04a} has reconsidered Einstein's ideas
in the context of the Pioneer anomaly and collected several
citations of Einstein that put into evidence that the principle of
constancy (over spacetime) of the speed of light is not a
necessary consequence of the principle of relativity. 
Furthermore, the notion of VSL is implicitly present in GR 
(see \cite{Bro:04}, ref.~70, citing numerous examples of GR textbooks). 
Despite this, there are physicists who consider a variable $c$ to be in
contradiction with pure logic, since the value of $c$ is defined
by the unit system. If however time and length scales which are
defined by atomic transitions
change accordingly, the change in $c$ is hidden at first glance. 
A drift of fundamental constants seems to provoke similar reactions
like a motion of the earth some hundreds of years ago:

`If earth is moving around the sun, why isn't there a strong wind blowing due to that
motion~?' was an argument of the adherents
of Ptolemy. Galilei responded:
\bq
`Close yourself with a friend in a possibly large room  below deck
in a big ship. [...] ...of all appearances you will not be able to deduce
a minute deviation...' \cite{Sin}
\eq
Sometimes it's useful to read Galilei, and sometimes it's useful to read Einstein.

\paragraph{Dirac's large number hypothesis,} like Mach's principle, is another
example of profound thoughts standing quite isolated
in the current fashion
of theories. Dirac \cite{Dir:38a} observed that all measured dimensionless quantities
in physics are either in the order of unity, or in the order of $10^{40}$ or $10^{80}$.
Suggesting that this can hardly be coincidence, he was speculating if the number of
protons in the universe is related to the square of the ratio of electric and
gravitational forces. Another example is  that the Hubble time $13.4 \ Gyr$ is about
$10^{40}$ times the time light needs to pass the proton radius. Though Dirac postulated
a time-dependence of the gravitational constant $\dot G/G$ which is above the current
observational constraints \cite{Uza}, these deep ideas should not be discarded
completely \cite{unz:07c}.

The wonderful `conceptual'  agreement of the quantum vacuum density and the cosmological
$\Lambda$, failing by just 120 orders of magnitude, is at least another interesting
number fitting into Dirac's considerations. Astonishingly,
field theorists usually do not appreciate Dirac's ideas.

\paragraph{Robert Dicke} is famous for his contribution to the CMB discovery and
for creating an alternative theory of gravity. This so-called Brans-Dicke or
scalar-tensor theory is in the meantime ruled out or at least reduced meaningless
by the $\om > 500$ constraints \cite{Rae:79}. Dicke's initial thoughts \cite{Dic:57} which
 have been anticipated by Sciama \cite{Sci:53}, however were
 much more general than the final form in collaboration with Brans \cite{Bra:61}.
Scalar-tensor theory sounds like an additional feature of GR, i.e. a potentially
unnecessary
complication. Dicke's original idea was similar to a pure scalar theory with
a variable speed of light. Here,  `scalar' does not mean a scalar 
theory coupled to matter to which later Einstein had expressed caveats in 
general.\fo{\cite{Giu:06} has given clarifying comments on that topic. It is pointed out that 
there is no shortcut as a matter of principle, rather
at the moment no convincing example of such a theory that matches the data.}

\paragraph{Anything else~?} In the outlooks or conclusions of modern textbooks
one usually can find the statement that after all, the standard model is the most 
likely option, after having mentioned MOND and its problems. But apart from being the most likely,
is it likely at all that our understanding is quite complete~? Why are people
so sure about this~? 
In his excellent book on the worrying situation of theoretical physics \cite{Smo:06}, Lee
Smolin gives an example of the mechanism that leads to such a narrowing of our view:
`The [...] possibility - that we are wrong about Newton's laws, 
and by extension general relativity - is too scary to contemplate.' (p.15).

\subsection{Basic problems}

Rather than postulating new quantities without meaning, advance in theoretical
physics has often been achieved
by a better understanding of quantities one was
already familiar with. There are still some points left where our knowledge must be scrutinized.
Research has shifted too much towards the technical issues, leaving behind the
fundamental questions.

\paragraph{Time.}
Barbour \cite{Bar} has presented a very deep reflection on the nature of time.
Analogously to the concept of space which does not make sense without matter,
he claimed that the concept of a time as an invisible river that runs without
relation to matter is senseless. If instead the evolution of the universe and
the periodic processes in there define time, profound consequences for
the laws of nature must be expected.

\paragraph{Energy.}
The concepts of potential and kinetic energy were born when
physicists described Galileo's free fall experiments with the
time-independent quantity \be m g h +\frac{1}{2} mv^2 = const. \ee
Newton's potential $GM/r$ and all the other forms of energy found
in the following were based on the same idea of finding
time-independent laws of nature. In quantum mechanics, this is
reflected by the fact that only stationary states of the wave
function have a well-defined energy. While the concept of energy
conservation is extremely successful in describing local effects,
its application to cosmology remains questionable. The
observations of the evolution of the universe, for instance the
CMB, the galaxy distribution and star formation processes tell us
that the wave function of the universe is anything but stationary.
Though standard cosmology (Friedmann-Lemaitre) is based on it, the
notions of kinetic and potential energy may be inadequate.\fo{ The
commonly used Euler-Lagrange formalism is just a general
mathematical tool on top.}
Is there a reason why energy should exist in two disperate and independent forms~?
Their separate existence is the reason why pure inertial motion is allowed
in Newtonian mechanics \cite{Bar:02}, something which is not really tested 
for small accelerations. The considerations
of Anderson \cite{And:06} are interesting in this context.

\paragraph{Mass.} is another poorly understood concept in physics. Barbour \cite{Bar:02}
has tried a  definition by means of inverse accelerations, but
what is behind that~? Einstein's famous $E= mc^2$, applied to
gravitational energy $\frac{G M m}{r}$, tells us that mass can be
proportional to a product of masses. This remains a very strange
fact, or could indicate that $G$ is an artefact that can be
calculated from parameters of the universe.

\paragraph{The equivalence principle,} the fact that a kinematic property measured
by accelerations, mass, at the same time should be  a `charge' for a certain interaction,
is very deep and puzzling. `I was ultimately astonished by its validity',
Einstein said \cite{EinstWB}. The (weak) EP is probably the most important constraint for
developing alternative theories, and its validity guarantees a special role for gravity
among the fundamental interactions.

\section{Methods}

\paragraph{A hierarchy of theory quality.}

\bq
{\it `correct theories of physics are perfect things, and a replacement theory 
must be a new perfect thing, not an imperfection added onto an old perfect thing. 
This is the essence of "revolution", the replacement of the old with a new, 
not the adding of more crap onto the old.' - Richard Feynman} 
\eq

The history of science hows many examples were knowledge
has not increased in a direct and steady way. Really fundamental theories
like electrodynamics and quantum mechanics yield revolutionary new insight
by reducing the number of the constants of nature, i.e. by the relation
$ \epsilon_0 \mu_0 =1/c^2$ or by expressing Rydberg's constant as a function
of $h, e, \epsilon_0$ and $m_e$. The next class consists of theories like QED or GR which
make testable predictions of extraordinary accuracy, such as for the Lamb shift
and the anomalous magnetic moment of the electron, and the classical tests of GR.
They `fail' however to derive quantities like $G$ and $m_e$, which is not excluded
as a matter of principle - in the case of $m_e$ this was considered by 
Feynman\footnote{See his Nobel lecture at $http://nobelprize.org/$.}, while 
Mach suggested that $G$ could be calculated from the mass distribution of 
the universe. The modern devolpments in theororetical physics are much less
ambitious, which is reflected by the term `model' rather than `theory'. 
The predictions merely regard the existence of particles or concepts, while 
quantitative predictions are abandoned by  introducting free parameters.
String and superstring `theories' have gone down further and
predict nothing - though in a self-confident manner \cite{Smo:06, Woi:06}.
Cosmology should be kept away from such a development.

\paragraph{Theories on top of theories.}
\bq
{\it `They defend the old theories by complicating things to the point of 
incomprehensibility.' - Fred Hoyle} 
\eq

Theoretical, as well as observational insight has hierarchical
structure. The research on reaction rates of complex molecules
would be on sandy grounds if there was any doubt on Mendelejew's
 order of chemical elements. Thus one has to be careful when
acquiring new knowledge on the basis of older one. This holds in
particular for astrophysics and cosmology, where no direct
experiments are possible. For instance, a different population of
cepheid stars, as initially used by Hubble, led to wrong estimates
of the age of the universe. Even every accepted theory may be
wrong with a small probability, and thus one has to take care of
the cumulating effect when building theories on top of others.

It is dangerous in particular when observational results are
interpreted with respect to a wrong theory. Almost five years of
research in nuclear physics were lost with a wrong theory of
transuranic elements, and the discovery of nuclear fission,
compared to the discovery of the cosmos, was an easy task. In
astrophysics, research has shifted its focus to galactic and
extragalactic observations, the regime in which the underlying
theory is least tested. While parameter measuring  becomes more
accurate, the interest as to the nature of DM seems to decrease
(while theoretical proposals for the recently discovered DE have
reached their peak). There is danger of brushing aside such
conceptional questions and of getting used to sandy grounds.

\paragraph{High-energy and particle physics.}
\bq
{\it `If you can't talk about something you have to hush.' - Ludwig Wittgenstein} 
\eq
The closer cosmology approaches the big bang, the more increases
the need to introduce knowledge from the smallest scales. Though
particle physics is not commented here\fo{The interested reader
is referred to Smolin's remarkable book \cite{Smo:06}.}, one can say that cosmology
did not develop  the first standard model with an inflation of
free parameters and  starry-eyed extrapolations of partly successful
theories have already occurred in physics.
 Sometimes one can hear that inflationary and particle models have
so similar concepts that there must be a deep relation.
There is a relation indeed: the closer we go to the big bang,
the more we can say that ignorance meets ignorance.

\paragraph{Separate data from interpretation.}
\bq
{\it `No theory should fit all data, because some data are surely wrong'.}
\eq
One possibility to do clean science is to separate observations
from theoretical interpretation, even if this might seem
complicated. Good science titles `Excess antenna temperature
observed at 4080 $Mc/s$'\fo{Original title of the CMB discovery.}
and not `God plays accordion'. 
 A tiny progress is
`cosmological redshift' instead of `Hubble expansion', and `faint
high-redshift SNIa' tells more than `dark energy measurement'.
Even if there is little doubt on a certain theoretical model, new
results must remain readable for those who
think about alternatives. 

Looking at alternative proposals, it is suspicious when a
surprising observational result  immediately backs a new theory at
hand. Even for true and honest scientists it makes sense to
maintain a distance between their theories and the experiments
supporting them. A famous example are the erroneous results of
Dicke and Goldenberg on the quadrupole moment of the sun that
should give evidence to a scalar-tensor theory of gravity.
Extraordinary claims need extraordinary evidence.

\paragraph{Science is quantitative.}

There are lots of results that cannot be understood with
conventional physics. With `conventional physics' I mean here
baryonic matter of known particles. It is not really scientific
to postulate unknown forms of matter if these forms for a long time
fail to show up in laboratory experiments. Moreover, is has become a habit
just to state that something does not work without dark matter. But how much~? There
are ratios for dark and visible matter for spiral and elliptic galaxies,
for clusters, superclusters, and there are estimates for DM needed
to explain structure formation, WMAP data, big bang nucleosynthesis,
and so on. Does all this really fit together in a quantitative way~?
It seems that every field of research fixes its problems with DM,
but hardly an encompassing calculation is done \cite{McG:07}.
To be concrete, the current parameter ratio $\Om_b, \Om_{DM}, \Om_{DE}$
is $0.02  :  0.24  : 0.74$.  How does this fit to ratios dark to visible
matter in the range of 100, for which we have evidence on
the galactic scale~? Where has all the DM gone since the discovery of DE~?

The use of numerical models, though useful for certain
 problems, is questionable whenever free parameters are introduced. Regarding the simulations of
structure formation, `Don't-the-pictures-look-alike~?' -results do not tell  whether
the underlying equations are correct.

\paragraph{Repeatability and source code.}

\bq
{\it `What is the chance that a person who notices an important discrepancy in a scientific announcement has the 
opportunity to check it out at the level of the primary data~?' - Halton Arp} 
\eq

Repeatability of results is a basic element of science. An
increasing number of results require extensive data reduction and
numerical calculations. It is obviously unsatisfactory if just one
group in the world can obtain a particular result (think about to
who it concerns), and oligarchy is only a partly solution. A
satisfactory solution would be if important results can be
verified by an unlimited number of scientists who have access to a
concise source code written in a widespread language or a computer
algebra system. As far as possible, access to the raw data should
be possible. An important step in this direction is the worldwide
availability of satellite data, the SDSS, or NASA's ephemeris site HORIZONS,
though the latter has a kind of monopole.

\paragraph{Unexpected is better than sought.}
Regarding our recent paradigm that includes `dark energy', it should
not be forgotten that without that interpretation, $H_0$ measurements were
in conflict with the age of the universe.
The revolutions in physics however did not start with looked-for
results but with completely unexpected ones: Michelson-Morley,
Perihelion advance, spectral lines of atoms, blackbody spectrum,
cosmic microwave background.\fo{Though this was predicted earlier.}
The Pioneer anomaly could fit into this line.


\paragraph{Generic and special} are terms of mathematical logic. To say that two planes
in threedimensional space do intersect and form a straight line,
is generically correct though there is the special case of
parallelism.\fo{Generically correct can be defined as correct
besides on a null set.} In astrophysics and cosmology, a lot of
observations seem to describe special cases (flatness,
coincidence, the `empty universe' at SNIa data etc.) while the
theory allows the general case. From a scientific point of view,
this is highly unsatisfactory, and `explaining' an apparent
parallelism of planes with a mechanism that decreases the
intersecting angle, as inflation does, is painting over rust.


\paragraph{Heisenberg, not Dirac.}
In Heisenberg's autobiography \cite{Hei} he reports on a
discussion with Dirac on the best method to attack physical
problems. Using the metaphor how to climb a steep mountain,
Dirac's opinion was to proceed step-by-step, since looking at the
whole task would be discouraging.  Heisenberg instead favored
planning the route strategically, since otherwise the risk of
getting into a dead end after following a promising track was too
high. To decide which way to take is not a scientific question,
but to get lost in details seems to be an obvious danger for
modern theories, too.
If Newtonian gravity turns out to be wrong, it will be not
an easy task to replace it. Such attempts should not try to
describe isolated phenomena but should at least have the potential to
encompass a wide range of anomalies. Given that it took Einstein about eight years
to develop general relativity \cite{Foe}, and a substantial modification of it 
is not supposed to be easier, the number of alternatives published every 
month is remarkable. People work fast today.

\paragraph{Publish or perish.}

\bq 
{\it `Do you know that the journal ... is going to be published 
with superluminal velocity~?' - `well, it does not contain information.' }
\eq

There is a tendency for publishing sensational discoveries, but a
lack of sustained systematic and methodic research projects.
\bq `... in the scientific magazines you can find lots of articles
on brilliant discoveries of galaxies, black holes or similar, but
you'll never find anything on those unsung heros  who for five
years were trying to improve our knowledge of opacity 
in
a dissertation 
without which we would never understand the internal processes in stars.'
\eq
(James Kaler, retranslation \cite{Kal}).

Many scientists mention the difficulty to publish non-mainstream results (e.g. 
\cite{Arp, Smo:06, Mag}). It seems that there is more danger in suppressing new 
unconventional ideas than in distracting scientists by exposing them
to speculative and potentially wrong papers. Reasons in the organization of science
cannot be discussed here, but some unsparing comments are given in \cite{LoC:08}.

\paragraph{Einstein and de Haas} measured 
the effect named after them in 1915, expecting a result of 1 for the gyromagnetic
relation. Experimentally they obtained 1.02 and 1.45, and discarded the latter
value because they thought it contained a systematic error \cite{Foe}. The published 1.02
was in `excellent agreement' with theory until the correct theory predicted 2
which was measured by others who repeated the experiment. This little story
tells us not to underestimate psychology, in particular because
(1) compared to current astrophysical observations, the Einstein-de Haas experiment
was a quite simple one
(2) Einstein was an open-minded, true researcher
(3) he did not need to become famous any more.

\section{Outlook}

\bq
{\it `The frog at the bottom of a well measures the extension of the sky with respect
to the border of the well.' (Chinese proverb)}
\eq

The theory of gravitation began with Galileo's $F=m g$, describing
effects on earth precisely. Newton's generalization of this formula and
the description of the solar system celestial mechanics  was an ingenious
big leap for science which at that time required an amount of
mathematical abstraction and new physical concepts we hardly
can imagine today. This holds even more for the refinement revealed by the general
theory of relativity.
Since those theoretical developments our knowledge of the universe has
increased drastically,
 in a relatively short span
of time: the discovery of galaxies, the Hubble redshift, the cosmic microwave
background. Due to satellite technique, improved telescopes and the computer-induced
revolution in image processing we are collecting data of fantastic quality
which allow to do quantitative cosmology for the first time.

I fear however that the hierarchy of structures earth - solar system -
galaxy - cosmos does not stop at laws found for the solar system
but requires a corresponding hierarchy of theories which may be similarly
hard to imagine like Newton's theory in 1600.
Introducing new parameters, data fitting and numerical simulations will
not do the job; rather this seems to be a modern version of
the deferrents, excentrics and equants
of the epicycles of Ptolemy.
It is not only the complication of our current theories that
merits a warning from history,
but also the amount of extrapolation we perform.
Though the range of the universe we know about has increased dramatically,
we  extrapolate conventional theories of gravity to those scales.
The extrapolation of classical
mechanics over 10 orders of magnitude to the atomic level was a
quite childish attempt, driven by the haughty attitude
of the end of the 19th century that the basic laws of physics were found
and only `corrections on the 6th decimal place' (Michelson) were needed.
Actually, this was the age of the first `standard model' of physics.
There were just two clouds on the horizon of 19th century physics:
the outcome of the Michelson experiment and the blackbody radiation.
Two thunderstorms came up, relativity and quantum mechanics.
The amount of research in cosmology which is done nowadays
on the base of an untested extrapolation over 14 orders of magnitude
is a quite remarkable phenomenon. It could be a good idea to pay
attention to the clouds. 
The wonderful observational data of the present should not lead us
to neglect the theoretical efforts of scientists in the past.
Sometimes one can learn more from the `blunders' of deep thinkers
like Einstein, Lord Kelvin, Mach or Sciama than from the latest theories in fashion.

\paragraph{Acknowledgement.}
The author thanks for discussions with Karl Fabian, Hannes Hoff and Jan Bernkopf, and
for the encouraging comments by gr-qc readers.

\end{document}